\documentclass[prl,showpacs,twocolumn,amsmath,amssymb]{revtex4}

\usepackage{graphicx}
\usepackage{dcolumn}
\usepackage{bm}

\begin{document}

\title{Free energy and torque for  superconductors with different
anisotropies of $\bm H_{c2}$ and $\bm \lambda$
}
\author{ V. G. Kogan }
\affiliation{ Ames Laboratory DOE and Department of Physics and
Astronomy, ISU, Ames, IA 50011, USA.}         
 
\date{July 29,2002}

\begin{abstract}
The free energy is evaluated for a uniaxial superconductor
with the  anisotropy of the upper critical field, $\gamma_H=
H_{c2,ab}/H_{c2,c}$, different from the anisotropy of the penetration
depth $\gamma_{\lambda}=\lambda_{c}/\lambda_{ab}$. With
increasing difference between $\gamma_H$ and $\gamma_{\lambda}$, the
equilibrium orientation of the crystal relative to the applied field
may shift from $\theta=\pi/2$ ($\theta$ is the angle between the
field and the $c$ axis) to lower angles and reach $\theta=0$ for
large enough $\gamma_H$. These effects are expected to take place in
MgB$_2$.
 \end{abstract}

\maketitle

It is a common practice to characterize anisotropic
superconductors   by a single anisotropy parameter defined as
$\gamma=\xi_a/\xi_c\equiv\lambda_c/\lambda_a$ ($\xi$ is the coherence
length, $\lambda$ is the penetration depth, and $a,c$ are principal
directions of a uniaxial crystal of the interest here). The practice
emerged after the anisotropic Ginzburg-Landau (GL) equations were 
derived for  arbitrary gap and Fermi surface anisotropies by
 Gor'kov and  Melik-Barkhudarov \cite{Gorkov}.   Formally, this came
out because in the GL domain, the same ``mass  tensor"   determines
the anisotropy  of both $\xi$ (of the upper critical fields
$H_{c2}$) and of $\lambda$.   

At arbitrary temperatures, however, the theoretical approach  for
calculating $H_{c2}$ (the position of the second order phase
transition in high fields) has little in common with evaluation of   
$\lambda$ (the weak-field relation between the
current and the vector potential), so that the anisotropies of these
quantities  are not necessarily the same. In fact, in materials with
anisotropic Fermi surfaces and anisotropic gaps, not only the
parameter $\gamma_H=H_{c2,a}/H_{c2,c}$ may strongly depend on  
$T$, but this ratio might differ considerably from
$\gamma_{\lambda}=\lambda_c/\lambda_a$ at low $T$'s. For MgB$_2$  
$\gamma_H\approx 6$ at low temperatures
\cite{BC,Angst,Welp,Weber}. There is no consensus yet on the low-$T$ 
$\gamma_{\lambda}$. Zehetmayer {\t et al} use the torque data to
find $\gamma_{\lambda}\approx \gamma_H$; however, their procedure
  implies this equality to begin with (as discussed in detail
below). Microscopic calculation for clean MgB$_2$ yields
$\gamma_{\lambda}\approx 1.1$  \cite{k,mmk}. With
increasing $T$, $\gamma_H(T)$ decreases, the calculated 
$\gamma_{\lambda}(T)$ increases till they meet at $T=T_c$: $\gamma_H
(T_c)=\gamma_{\lambda}(T_c)\approx 2.6\,$.

One of the most sensitive methods used to extract the anisotropy
parameter $\gamma$ is to measure the torque acting on a
superconducting crystal in the mixed state situated in the applied
field tilted relative to the crystal axes. In intermediate field
domain, $H_{c1}\ll H\ll H_{c2}$, the demagnetization shape effects
are weak, and the torque density can be evaluated \cite{K88,farr}: 
\begin{equation}
\tau = \frac{\phi_0B(\gamma^2-1)\sin
2\theta}{64\pi^2\lambda^2\gamma^{1/3}
\varepsilon(\theta)}\, \ln\frac{\eta\,H_{c2,a} } { B
\varepsilon(\theta) } \,, 
\label{trq_0}
\end{equation}
where $\theta$ is the angle between the induction $\bf B$ and the
crystal axis $c$, 
\begin{equation} 
\varepsilon(\theta) =\sqrt{\sin^2\theta+\gamma^2 \cos^2\theta}\,,
\label{eps}
\end{equation}
$\lambda^3=\lambda_{ab}^2\lambda_c$, and $\eta\sim 1$. This formula
can be written as ${\bm \tau}={\bm M}\times{\bm H}$; since in this
field domain the magnetization $M\ll H$, one can disregard the
distinction between $B$ and $H$ so long as the torque is concerned.
It has been assumed in the derivation of Eq. (\ref{trq_0})  that the
anisotropies of $H_{c2}$ and of the London penetration depth are the
same: $\gamma_H =\gamma_{\lambda} =\gamma$. 

Note that Eq. (\ref{trq_0}) describes the system with the stable
equilibrium at $\theta=\pi/2$, i.e., the uniaxial crystal in the
external field  positions itself so that the   field is parallel to
$ab$ (if $\gamma > 1$). It is also worth noting that if one applies
formally the expression (\ref{trq_0}) with 
$H_{c2} (\theta)=H_{c2,a}/\varepsilon(\theta)$ having the
anisotropy different from that of $\lambda$, one may obtain
qualitatively different angular behavior of $\tau (\theta)$. 

Expression (\ref{trq_0}) for the torque has been derived within the
London approach by employing the cutoff at distances on the order of
the coherence length $\xi$ where this approach fails; that
is how the upper critical field $H_{c2}\sim \phi_0/\xi^2$ enters  
the London formula. The formula, however, has been confirmed
experimentally with a good accuracy as far as the angular dependencies
of quantities involved are concerned \cite{Far2}. Uncertainties of the
London approach are incorporated in the parameter $\eta\sim 1$;
discussion of those can be found, e.g., in Ref. \onlinecite{nonloc}. 

Below, the free energy of the mixed state and the torque are
 derived for the general case of $\gamma_H \ne\gamma_{\lambda}$. The
torque expression is shown to acquire new terms which describe a more
complicated behavior as compared to that of Eq. (\ref{trq_0}). This
implies that the analysis of the torque data might be misleading if
Eq. (\ref{trq_0}) is employed to materials like
MgB$_2$ \cite{torque1,Weber,torque2}.\\

 
Let us start with the London expression for the free energy valid for
intermediate fields $H_{c1}\ll H\ll H_{c2}$   along $z$ 
tilted with respect to the $c$ crystal axis over the angle $\theta$
toward the crystal direction $a$ \cite{CDK}:
\begin{equation}
F=\frac{B^2}{8\pi }+\frac{B^2m_{zz}}{8\pi\lambda^2m_a}\, \sum_{\bm
G}\frac{1}{m_{zz}G_x^2+m_cG_y^2}\,;
\label{energy}
\end{equation}
here, $m_{zz}=m_a\sin^2\theta+m_c\cos^2\theta $,
$m_c/m_a=\gamma_{\lambda}^2\,$, $m_a^2m_c=1$ for uniaxial crystals, and
${\bm G}$ form the reciprocal vortex lattice. The summation is
extended over all nonzero ${\bm G}$. 

As usual we evaluate the sum here by replacing it with an integral
over the reciprocal plane: $\sum_{\bm G}\to (\phi_0/4\pi^2
B)\int dG_xdG_y\, $:
\begin{eqnarray}
{\tilde F}=F-\frac{B^2}{8\pi
}&=&\frac{\phi_0Bm_{zz}}{32\pi^3\lambda^2m_a}\,
\int \frac{dG_xdG_y} {m_{zz}G_x^2+m_cG_y^2}\,\nonumber\\
&=&\frac{\phi_0B\sqrt{m_{zz}}}{32\pi^3\lambda^2
}\,\int_0^{2\pi}d\varphi
\int \frac{dg\,} {g}\,,
\label{F'}
\end{eqnarray}
where  $g_x=\sqrt{m_{zz}}\,G_x\,$, $g_y=\sqrt{m_c}\,G_y\,$, and we use
polar coordinates: $g_x=g\sin\varphi, g_y=g\cos\varphi$. 

To determine the limits of integration over $g$, one can recall that
in fields $B\ll H_{c2}$ the vortex lattice structure is fixed by
the anisotropy parameter $\gamma_{\lambda}$ \cite{CDK}. In
tintermediate fields, for all possible equilibrium London structures,
the distance in the reciprocal space ${\bm g}$ from the origin to the
nearest neighbors is given by \cite{rem1}
\begin{equation}
g_0^2=m_{zz}G_x^2+m_cG_y^2 =
\frac{8\pi^2B}{\sqrt{3}\phi_0}\sqrt{m_c m_{zz}}\,.
\label{go}
\end{equation}
Therefore, one can take $g_0$ (multiplied by a number of order
unity) as the lower limit in the logarithmically divergent integral
over $g$. 

The upper limit in this integral  
is affected by the form of the vortex core. To determine the core
shape we note the microscopic evaluation of $ H_{c2}$ (see Ref.
\onlinecite{mmk}) shows that with a good accuracy the angular
dependence of $H_{c2}$ is given by the standard GL form at any $T$:
\begin{equation}
H_{c2}=\frac{\phi_0}{2\pi\xi^2\sqrt{
\mu_a\sin\theta+\mu_c\cos\theta}}=
\frac{\phi_0}{2\pi\xi^2 \sqrt{\mu_{zz}}}\,.
\label{Hc2}
\end{equation}
This can be written as 
\begin{equation}
H_{c2}=\frac{\phi_0}{2\pi\xi_x\xi_y }\,,\quad
\xi_x=\xi\sqrt{\mu_a\mu_{zz}}\,,\quad
\xi_y={\xi\over \sqrt{\mu_a }}\,,
\label{xi's}
\end{equation}
where $\xi_{x,y}$ are semiaxes of the elliptical core. We stress that
the ``masses" $\mu_{ik}$ are different from   $m_{ik}$ which
determine the anisotropy of $\lambda$; in particular,
$\mu_a=\gamma_H^{-2/3},\quad \mu_c=\gamma_H^{4/3}$, whereas
$m_a=\gamma_{\lambda}^{-2/3},\quad m_c=\gamma_{\lambda}^{4/3}$.
Equation (\ref{xi's}) gives  maximum values of $G_x$ and $G_y$:
\begin{equation}
G_{x,m}=\frac{2\pi}{\xi\sqrt{\mu_a\mu_{zz}}}\,,\quad
G_{y,m}=\frac{2\pi\sqrt{\mu_a }}{\xi}\,.
\label{Gmax}
\end{equation}
Thus, the domain of integration in the ${\bm G}$ plane is bound  by
an ellipse
\begin{equation}
\frac{G_x^2}{G_{x,m}^2} + \frac{G_y^2}{G_{y,m}^2}=1\,.
\label{ellipse}
\end{equation}
In other words, at a given $\varphi$, the upper limit in the integral
over $g$ is 
\begin{equation}
g_m(\varphi) =
\frac{2\pi\sqrt{\mu_am_c}}{\xi\sqrt{\beta^2\cos^2\varphi
+\sin^2\varphi}}\,,\quad \beta^2=\frac{m_c\mu_{zz}}{m_{zz}\mu_c}\,,
\end{equation}
and we obtain:
\begin{eqnarray}
{\tilde F}&=& \frac{\phi_0B\sqrt{m_{zz}}}{32\pi^3\lambda^2
}\,\int_0^{2\pi}d\varphi\, \ln\frac{g_m(\varphi)}{g_0}\nonumber\\
&=&\frac{\phi_0B\sqrt{m_{zz}}}{32\pi^2\lambda^2}\,\Big[\ln\frac{
\sqrt{3m_c}\,\mu_a\phi_0}{2\xi^2B\sqrt{m_{zz}}}\nonumber\\
&-&{1\over 2\pi}
\int_0^{2\pi}d\varphi\, \ln(\beta^2\cos^2\varphi
+\sin^2\varphi)\Big]\,. \label{tilde_F}
\end{eqnarray}

Note that   $\beta=1$ for  coinciding anisotropies of
$H_{c2}$ and $\lambda$; the integral $J(\beta)$ over $\varphi$ then
vanishes and we have the standard expression for the energy. To
evaluate  $J(\beta)$, we observe that the integral for $dJ/d\beta$  
 is a rational function of $\cos^2\varphi$  and therefore   can be
calculated by going to the complex plane with the help of  residues:
$dJ/d\beta=4\pi/(\beta+1)$. Since $J(1)=0$ we obtain:  
\begin{equation}
J =4\pi\,\ln {1+\beta\over 2}\,. 
\end{equation}
Thus, we have 
\begin{equation}
{\tilde F}= \frac{\phi_0B\sqrt{m_{zz}}}{32\pi^2\lambda^2}\,
 \ln\frac{2
\sqrt{3m_c}\,\mu_a\phi_0}{\xi^2B\sqrt{m_{zz}}\,(1+\beta)^2} \,. 
\label{F(theta)}
\end{equation}

To write explicitly the angular dependence of $F$, it is convenient to
use the angular functions 
\begin{equation}
\Theta_{\lambda,H}(\theta)=\varepsilon_{\lambda,H}(\theta)/\gamma_{\lambda,H} 
\end{equation}
where   $\varepsilon_{\lambda,H}(\theta)$ is defined in Eq.
(\ref{eps}) with corresponding $\gamma$'s.
In terms of these functions, $\beta=\Theta_{ H}/\Theta_{\lambda
}$ and 
\begin{equation}
{\tilde F}= \frac{\phi_0B \Theta_{\lambda
}}{32\pi^2\lambda_{ab}^2}\,
 \ln\frac{2
\sqrt{3}\,\mu_a\phi_0\Theta_{\lambda
}}{\xi^2B \,(\Theta_{\lambda }+\Theta_{H })^2} \,. 
\label{FF}
\end{equation}\\


The torque density follows:
\begin{eqnarray}
\tau=-{\partial \tilde F\over \partial\theta}=&-&
\frac{\phi_0B
}{32\pi^2\lambda_{ab}^2}\,\Big[\Theta_{\lambda}^{\prime}
\, \ln\frac{2e
\sqrt{3}\,\mu_a\phi_0\Theta_{\lambda
}}{\xi^2B \,(\Theta_{\lambda }+\Theta_{H })^2}\nonumber\\
&-&2 \Theta_{\lambda}
\frac{\Theta_{\lambda}^{\prime}+\Theta_H^{\prime}}
{\Theta_{\lambda}+\Theta_H}\Big]\,, 
\label{trq}
\end{eqnarray}
where $e=2.718...\,$.
The torque is zero at $\theta=0,\pi/2$ because  
\begin{equation}
\Theta ^{\prime}
=-\frac{(\gamma ^2-1)\,\sin
2\theta}{2\,\gamma ^2\,\Theta }
 \,. \label{e'} 
\end{equation}
for both $\Theta_{\lambda}$ and $\Theta_H$. 
In the standard case of $\gamma_H=\gamma_{\lambda}=\gamma$, Eq.
(\ref{trq}) reduces to the  result (\ref{trq_0}) if we set
$\eta=\pi\sqrt{3}/e\approx 2\,$. We then can rewrite the torque in
the form:
\begin{eqnarray}
\tau  &=&
\frac{\phi_0B(\gamma_{\lambda}^2-1)\sin 2\theta}
{64\pi^2\lambda^2\gamma_{\lambda}^{4/3}\Theta_{\lambda}}\,
\Big[\ln\Big(\frac{\eta H_{c2,c}}{B}\,\frac{4e^2
\Theta_{\lambda}
 }{(\Theta_{\lambda }+\Theta_{H })^2}\Big)\nonumber\\
&-&
\frac{2
\Theta_{\lambda}}
{\Theta_{\lambda}+\Theta_H}\Big(1+\frac{\Theta_H^{\prime}}
{\Theta_{\lambda}^{\prime}}\Big)\Big]\,. 
\label{trq1}
\end{eqnarray}

Since  $\Theta_{\lambda,H}^{\prime} < 0$, the
second  contribution to the torque (\ref{trq}) is negative whereas the
first one is positive. The positive torque implies that the system
energy decreases with increasing $\theta$, as in the case of
$\gamma_H=\gamma_{\lambda}$ for which   $\theta=\pi/2$ is the
stable equilibrium. 

The competing roles of this two contributions can be demonstrated by
considering stability of equilibrium states at
$\theta=0$ and $\theta=\pi/2$. To do this one notes that
$\Theta (0)=1$, $\Theta (\pi/2)=1/\gamma $, $\Theta
^{\prime}(0)=\Theta ^{\prime} (\pi/2)=0$, and 
\begin{equation}
\Theta^{''}(0)=-\frac{\gamma^2-1}
{\gamma}\,, 
\qquad \Theta^{''}(\pi/2)= \frac{\gamma^2-1}
{\gamma}\,
\end{equation}
(for both $\Theta_{\lambda}$ and $\Theta_H$).
 Then, one   obtains:
\begin{eqnarray}
F^{''}\Big({\pi\over 2}\Big)&= &
\frac{\gamma_{\lambda}^2-1}{\gamma_{\lambda}}\,\ln
\frac{4\pi e\sqrt{3}H_{c2,a}\gamma_{\lambda}\gamma_H}
{B(\gamma_H+\gamma_{\lambda})^2}
-2 \frac{\gamma_H \gamma_{\lambda}-1}
 {\gamma_{\lambda}} \,,\nonumber\\
F^{''}(0) &=& -
\frac{\gamma_{\lambda}^2-1}{\gamma_{\lambda}}\,\ln
\frac{\pi\sqrt{3}H_{c2,c}}{ B}
+ \frac{\gamma_H^2-1}{\gamma_H}\,,
\label{0-pi/2}
\end{eqnarray} 
where
 the   constant positive prefactor is  omitted since we are
interested only in the sign of $F^{''}$.  Clearly,   $\theta=\pi/2$
corresponds to the stable  equilibrium for $\gamma_H=\gamma_{\lambda}
$. In the general case, however, there is no such a clear-cut result:
for a fixed $\gamma_{\lambda}$ and large enough
$\gamma_H$, $\theta=\pi/2$ may become unstable. E.g., for
$\gamma_{\lambda}=1$ and $\gamma_H>1$, $F^{''}(\pi/2)<0$ whereas
$F^{''}(0)>0\,$. \\

\begin{figure}[t]
\includegraphics[angle=0,scale=0.6]{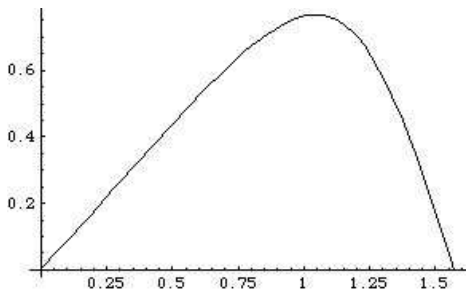}
\caption{The torque in units of
$\phi_0B /32\pi^2\lambda_{ab}^2$ {\it versus} angle
$0<\theta<\pi/2$ for $\gamma_{\lambda}=2.2$, $\gamma_H=3$, and
$4e^2\eta H_{c2,c} /B=100$. }
\label{fig1}
\end{figure}

To illustrate how the angular dependence of the torque varies with 
 anisotropies of $H_{c2}$ and  $\lambda$,  we evaluate numerically
the torque density (\ref{trq1}) for parameters in the range of those
for MgB$_2$.  Fig. \ref{fig1} shows $\tau(\theta)$ for
$\gamma_{\lambda}=2.2$ and $\gamma_H=3$, the values expected for
temperatures somewhat below $T_c$. Qualitatively, the
dependence is standard; the torque is positive in the whole angular
domain, i.e.,
$\theta=\pi/2$ is the stable equilibrium.  

\begin{figure}[t]
\includegraphics[angle=0,scale=0.6]{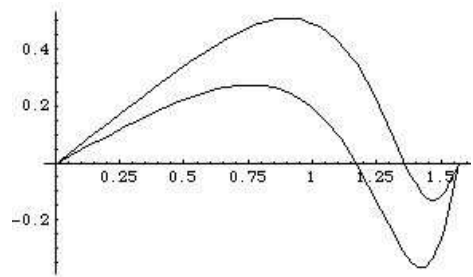}
\caption{The same as in Fig.\ref{fig1}. The
upper curve is calculated with Eq. (\ref{trq1}) for
$\gamma_{\lambda}=2$ and $\gamma_H=5$;  at the lower curve 
$\gamma_{\lambda}=1.7$ and $\gamma_H=5.3\,$.}
\label{fig2}
\end{figure}

With decreasing $T$, $\gamma_H$ of MgB$_2$ increases whereas 
$\gamma_{\lambda}$ decreases. In Fig. \ref{fig2} the torque
(\ref{trq1}) is plotted for $\gamma_{\lambda}=2$, $\gamma_H=5$
(the upper curve) and for $\gamma_{\lambda}=1.7$, $\gamma_H=5.3$ 
(the lower curve). These values 
correspond roughly to  0.7 and 0.6$\, T_c$ according to Ref.
\onlinecite{mmk}. Clearly, $\theta=\pi/2$ as well as
$\theta=0$ are unstable; the stable equilibrium is shifted to
$0<\theta<\pi/2$.  Interestingly enough,  Angst
{et al.} observe that a strong ``peak-effect-like" irreversibility
develops in the torque data  at
$\approx 77^{\circ}$ at $T=15\,$K and $H=7.5\,$T \cite{Angst2}.
In layered materials this effect is commonly interpreted as
manifestation of the ``intrinsic pinning" in the small vicinity of
the equilibrium orientation at $\theta=\pi/2$. From the point of view
of this article, the peak should move to a position of the stable
equilibrium, i.e., to lower angles. Moreover, the data show a {\it
negative} torque above this angle ($77^{\circ}<\theta <90^{\circ}$) in
agreement with Fig. \ref{fig2}.  Still, interpretation of these data
within the London model should not be taken too seriously:  the
applied field  $H=7.5\,$T exceeds $H_{c2,c}(0)$   and in the
$\theta$-domain where $H< H_{c2 }(\theta)$, the London model should
not be trusted. 

\begin{figure}[t]
\includegraphics[angle=0,scale=0.6]{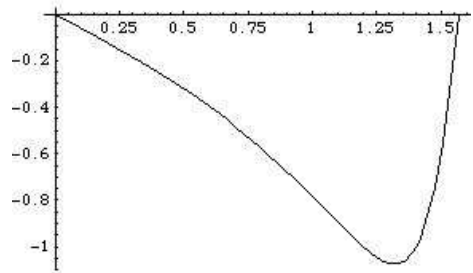}
\caption{The same as in Fig.\ref{fig1}, but $\gamma_{\lambda}=1.1$
and $\gamma_H=6$.}
\label{fig3}
\end{figure}

Finally, we plot in Fig. \ref{fig3} the torque density for parameters
which correspond to low temperatures, where 
$\gamma_{\lambda}\approx 1.1 $
and $\gamma_H\approx 6$. The torque is negative for all angles
implying  the stable equilibrium at   $\theta=0$.  

Physically, the large low temperature anisotropy of $H_{c2}$ in
MgB$_2$ is caused by the large superconducting gap on the nearly
two-dimensional sheets of the Fermi surface of this material
\cite{Bel,Mazin,Choi}. With increasing $T$, thermal mixing with the
states at the three-dimensional part of the Fermi surface suppresses
this anisotropy to about $\gamma_H(T_c)\approx 2.6$. The anisotropy
of the London $\lambda$ (or of the superfluid density) at $T=0$ of
clean materials does not depend on the gap at all (``Galilean
invariance of the superflow") and therefore is determined by the
whole Fermi surface, i.e., it is weak for MgB$_2$ (see  discussions
in Ref. \onlinecite{k,mmk}). The calculation \cite{k} shows that
$\gamma_{\lambda}(T=0)\approx 1.1$ and grows to $\approx$ 2.6 as
$T\to T_c$.

Thus, different gaps at different Fermi surface pieces (or  
generally,  anisotropic gaps on anisotropic Fermi surface) may lead
to profound macroscopic consequences such as those considered above.
Certainly, the equilibrium magnetization and the flux lattice
structure should be strongly affected by differences in anisotropies
of the upper critical field and of the London penetration depth.
Occurrence of a peak-effect near the field orientation other than
$H\parallel ab$ is an example of peculiar dynamic phenomena which
call for further study. 

I thank  P. Miranovi\'c  and M. Angst for  numerous discussions which
led me to consider the problem of this paper. 
  Ames Laboratory is operated for the U. S. Department of Energy by
Iowa State University under Contract No. W-7405-Eng-82. This work  
is supported by the Office of Basic Energy Sciences.


\begin{thebibliography}{99}
\bibitem{rem1} This result comes about because the vortex
lattice hexagon of the isotropic case have the same energy for all
possible orientations; the nearest neighbors are all situated at
a circle of a certain radius. The anisotropy amounts to rescaling of
$x$ and $y$ directions by different factors, i.e., transforms the
the nearest neighbor circle to an ellipse.  All possible structures
so obtained have the same energy in the approximation used here
\cite{CDK}.
\end{thebibliography}
\end{document}